\documentclass[dvips]{arxstspdf}
\usepackage{flushend}

\volume{24}
\issue{3}
\pubyear{2009}
\firstpage{270}
\lastpage{272}
\doi{10.1214/09-STS277A} \referstodoi{10.1214/09-STS277}

\begin{document}
\begin{frontmatter}
\vspace*{9pt}

\title{Discussion of Likelihood Inference for Models with
Unobservables: Another View}
\runtitle{Discussion}
\pdftitle{Discussion on Likelihood Inference for Models with
Unobservables: Another View by Y. Lee and J. A. Nelder}

\begin{aug}
\author[a]{\fnms{Thomas A.} \snm{Louis}\ead[label=e1]{tlouis@jhsph.edu}\corref{}}
\runauthor{T. A. Louis}

\affiliation{Johns Hopkins Bloombergh School of Public Health}

\address[a]{Thomas A. Louis is Professor, Department of Biostatistics,
Johns Hopkins Bloombergh School of Public Health, Baltimore, Maryland, USA \printead{e1}.}

\end{aug}

% ABSTRACT

% KEYWORDS

\end{frontmatter}

%s1 ###
\section{Introduction}

Lee and Nelder identify important issues and provide
excellent advice and warnings associated with inferences and
interpretations for models with unobserved, latent variables (random effects).
Their discussion of prediction versus estimation goals is insightful
and I~have some sympathy with their call for use of comprehensive
probability models. They provide a clear explanation of their
h-likelihood approach and a spirited promotion of it.
Unfortunately, the value and impact of the their advice are
compromised by their singular focus on promoting h-likelihood. Their
claim that it is an almost universally preferred approach is, to put it
mildly, a stretch.
The h-likelihood approach by no means ``trumps'' all competitors and has
its own deficits.
Over promotion makes the article more of an opinion-piece than a
scientific comparison of approaches.

%A more balanced view would be very informative and beneficial.
%It would have been great to have some additional concrete examples,
%not just formulations.

%s2 ###
\section{Point/Counterpoint}
I identify and discuss principal points of (partial) agreement and of
disagreement. Statements by Lee and Nelder are in \textit{italics}; my
responses and
comments are in Roman.

%s2.1 ###
\subsection{Modeling Strategies}
Lee and Nelder write, ``\textit{However, we believe that such a choice is
inappropriate because the choice of an estimation
method for a particular parameterization (marginal parameter) should
not pre-empt
the process of model selection.}'' I agree. Estimation methods are a
means to an end and usually not, themselves, the end (in methods
research they can be the goal). Of course, the estimation method might
influence model choice in that an inefficient method may miss important
covariates and an inappropriate method may lead to bias.
Sometimes the means/ends distinction gets blurred. For example, several
years ago someone wrote to let me know that he thought the EM estimate
was the absolute best; far better than the MLE!

\textit{Unified Probability Models are absolutely necessary:}
I do take issue with this claim. One should not discount the
effectiveness of analyses and algorithms that are not fully
probability-based or comprehensive. These have and will continue to
play an important role.
% with some ultimately being embedded in a probability model framework
%and others not being so embedded.
While a unified approach with marginals, conditionals, etc., all
generated by a joint distribution is without question the ideal, often
it is not attainable. Data limitations, limitations in scientific
understanding and computing constraints can thwart use of this holy
grail. Even attainment can be illusory because the unified model may
not be correct and may mislead.
So, while I favor the unified approach, I'm very comfortable with an
approach that validly and effectively addresses a specific goal.

``\textit{\ldots so that care is necessary in making inferences about
unobservables.}''  Absolutely! Extreme care and caution are most
definitely needed. Inferences on latent effects are always model-based
to some degree, and some assumptions cannot be verified empirically.
%Therefore, such inferences are inevitably based on some unverifiable
%assumptions.
%Use of the h-likelihood approach, doesn't avoid this problem.
For example, models using the standard Poisson distribution as baseline
rather than the more general negative binomial will ``identify'' unaccounted (extra-Poisson) variation and allocate it to a latent
effect. If a negative binomial model is used, much of this variation
will be absorbed into the baseline model. Both approaches can produce
similar predictions of observable quantities, but will produce very
different inference for latent effects.
All modeling approaches need to deal with such issues, and the
h-likelihood is not a panacea.
In contrast, use of latent variable models and hierarchical models to
generalize the mean and association structure of models for observeds
is quite safe. Therefore, I~agree with Lee and Nelder that focus on the
prediction space rather than the parameter space avoids mis- or
over-interpretation of parameter estimates.

%s2.2 ###
\subsection{H-likelihood and Competitors}

``\textit{\ldots that when applied appropriately h-likelihood\break
methods are both valid and efficient in such settings.}''  It is most
surely the case that in some settings, with an appropriate
parameterization, the h-likelihood approach is valid and efficient.
However, it is not globally valid and even when it is valid may perform
no better than, and possibly worse than other approaches.

``\textit{However, GEEs cannot (generally) be integrated to obtain a likelihood
function [McCullagh and Nelder (1989)] and therefore may not have a
probabilistic or
likelihood basis.}'' True, but GEEs can be very effective, especially
for population-targeted inferences.
I agree with Lee and Nelder that likelihood-based approaches or likelihood-like
(marginal, partial, profile,\ldots) approaches should be used when
available and their use is essentially necessary when making inferences
on latent effects.

``\textit{HGLMs allow a synthesis of GLMs, random-effect models, and
structured-dispersion
models.}'' They do synthesize, but aren't alone in accomplishing this task.

%{21:} You write, ``Without probabilistic models it is not possible to
%connect these parameters and compare them. We dislike the
%pre-emption of the model selection stage by a particular estimating
%procedure.'' Great point!

%%%%%%%%%%%%%%%%%%%%%%%%%%%%%%%%%%%%%%%%%

\textit{Bayes is like the Adjusted profile h-likelihood\break  (APHL).} Well,
that's one way to put it. The other way is that the APHL is like Bayes.
Regarding extended Likelihood versus (empirical) Bayesian approaches,
one can think of the h-likelihood as prior augmented likelihood, an
attractive approach to stabilizing and \mbox{smoothing} MLEs.
However, taking full advantage of the structure requires moving away
from mode/curvature inferences and, at least for some
nonstandard goals, \mbox{employing} the fully Bayesian formalism.

\subsubsection*{Poor performance of plug-in empirical Bayes (EB)}
Yes, naive EB
produces a too-low variance estimate (more generally, an incorrect
shape and association structure), unless the estimates of prior
parameters are very precise. This observation motivated the\break  Laird/Louis
bootstrap and Carlin and Gelfand's\break matching approach. These have been
supplanted by Bayes empirical Bayes (BEB) with a hyper-prior from which
prior parameters are sampled. BEB has proven very \mbox{effective} in
producing procedures with excellent frequentist (as well as Bayesian)
properties.
See, for \mbox{example}, Table~3.4 in \citet{carlloui2000} and Table~5.6 in
\citet{carlloui2009}.

%It is most definitely the case that more research is needed on
%selecting hyper-priors that produce good frequentist properties, but
%BEB is the right way to go.

\subsubsection*{Priors and hyper-priors}
 Lee and Nelder state, ``\textit{In Bayesian
analysis, priors can give information on unidentifiable model
assumptions, so that it is hard to know whether the information is
entirely coming from the uncheckable priors.}''
Yes, and ditto for modeling assumptions whatever the approach. Care is needed.

%; neither a Bayesian nor an h-likelihood approach is a panacea.

In Section~4.3.1, Lee and Nelder criticize use of $\sigma^{-2} \sim
\operatorname{gamma}(0.0001, 0.0001) $. The problems with using this prior and a
$\operatorname{gamma}(\alpha, \alpha)$ more generally are well known. Though the mean
is 1 and coefficient of variation is large, most of the prior mass is
in the interval $(0, 1]$. It's better to use a uniform prior on $\log
{\sigma}$ in a bounded interval with the bounds selected to respect
measurement units. It is most definitely the case that more research is
needed on selecting hyper-priors that produce good frequentist
properties. This and other examples highlight the need for
sophistication and care when exploring the latent world.

%s2.3 ###
\subsection{Goals that Challenge the H-likelihood}
%%%%%%%%%%%%%%%%%%%%%%
\subsubsection*{Accounting for uncertainty}
Lee and Nelder make the important point about the need to account for
uncertainty, but can't avoid ``dissing'' (empirical) Bayes. They state,
``\textit{The h-likelihood approach takes
into account the uncertainty in the estimation of random effects, so
that inferences about
unobservables are possible without resorting to an EB framework.}''  The
\mbox{h-likelihood} may take this uncertainty into account, but it does not
ensure that all relevant uncertainties migrate into the inferences. For
example, it does not allow for adjusting the shape of or association
structure in the distribution of random effects, whereas the Bayesian
formalism introduces both of these along expanding the spread by
integrating over the posterior hyper-prior.

\subsubsection*{Nonstandard goals}
Regarding goals, while the\break  \mbox{h-likelihood} and other purely
likelihood-based approaches can be effective in making inferences on
measures of central tendency and linear functions of \mbox{target} parameters,
they have a difficult time in structuring an approach for nonstandard
goals\break whereas the Bayesian formalism is successful. For example,
consider estimating the ranks of the $\theta_k$ in a two-stage model,
$[\theta_1, \ldots, \theta_K]$ i.i.d. $G; [Y_k \vert \theta_k]$ i.n.d. $f_k(y_k
\vert \theta_k)$. As detailed in \citet{linetal2006}, if the $\theta$s
were observed, $R_k(\bolds{\theta}) = \sum_{\nu= 0}^{K} I_{\{
\theta_k \ge
\theta_\nu\}}; P_k =\break R_k/(K+2)$ with the smallest $\theta$ having rank
1.\break Ranks/percentiles that minimize posterior expected squared-error
loss {for the ranks} are their posterior mean or a discretized version,
\begin{eqnarray*}
\bar{R}_k(\mathbf{Y}) & = & E[R_k(\bolds{\theta}) \mid \mathbf{Y}]
= \sum_\nu \operatorname{pr}[ \theta_k \ge\theta_\nu\mid\mathbf{Y}], \\
\hat R_k &=& \mbox{rank}(\bar{R});\qquad \hat P_k = \hat R_k/(K+1).
%= \sum_\nu P_{k \nu}\\[3ex]
\end{eqnarray*}
The model can be generalized to BEB and is effective in both Bayesian
and frequentist evaluations.
Similarly challenging inferential goals are handled well (if handled
with care!) by the Bayesian formalism, including proper accounting for
uncertainty.

%%%%%%%%%%%%%%%%%%%%%%%%%%%%%%%%%%%%%%%%%%%%%%%%%%%%%
\subsubsection*{Computational challenges}
Lee and Nelder write, ``\textit{However, the computation of the ML
estimation of the parameters can be a complex task because of
intractable integration.}''
Yes, finding the MLE and developing appropriate inferences can can be
complex, and expansions around the mode may not be up to the task.
Markov chain Monte Carlo methods have enabled likelihood-based and
Bayesian-based analyses of complex data and models. Use them, but carefully!

\subsubsection*{Resorting to (empirical) Bayes}
It is strange that Lee and Nelder characterize use of empirical Bayes a
``resort.'' In this day and age is the Bayesian formalism to be avoided?
Have the last 20--25 years passed Lee and Nelder by?
Most statisticians have gone beyond the Bayes/frequentist polemic of
the 1980s and early 1990s. Yes, there are challenges, but use of the
Bayesian formalism in both it's objective and informative-prior forms,
burgeons. It's use is by no means a panacea, but carefully employed, it
is very effective in addressing both Bayesian and frequentist goals.

% proven to be most effective in computing integrals, irrespective of
%whether your target is a posterior distribution,
%a likelihood or an h-likelihood. Expansions and approximations
%(Laplace) are not really competitive.
%Works well for Gaussian-like situations.

%You are really doing approximate likelihood inferences.
%You are doing Bayes with some approximations. Why not just so state
%and then study properties? You may find that in some %situations the
%shape of things via your method is poor. Give the C\&L EB example.

%%%%%%%%%%%%%%%%%%%%%%%%%%%%%%%%%%%
%s3 ###
\section{Summary}
Lee and Nelder provide considerable food for\break  thought, considerable
light and some
heat, heat produced by their over-promotion of h-likelihood.
%I have also pitched in with a bit of polemic.
I support Lee and Nelder's goal of attempting a unified analysis based
on full probability modeling, but note that the Bayesian formalism is
best suited to this task.\ Use of the full probability calculus,
empowered by modern computing, brings in (most) relevant uncertainties,
produces properly shaped and calibrated confidence regions and enables
addressing nonstandard goals such as ranking.
However, I caution that full probability modeling isn't always
available or valid and in many situations compromises are necessary.

%While Lee and Nelder's h-likelihood is effective in many situations,
%it is not always so. The same caution holds for all approaches.
% including use of the Bayesian formalism, but the latter is my
%preferred strategy. It includes likelihood-based inferences via
%``flat'' priors.

Whatever the approach to analysis, care, evaluation, and sophistication
are needed, especially when structuring inferences for latent effects.
Polemic and over-promotion distract from the important issues and
goals. These should be replaced by aggressive scientific evaluations
and energetic discourse.
%In that regard, Bayes-bashing must end and I encourage \len to engage
%in dispassionate comparisons and evaluations.

\section*{Acknowledgment}
Support provided by Grant R01 DK061662, from the U.S. NIH, National
Institute of Diabetes, Digestive and Kidney Diseases.

\vspace*{-2pt}
\end{document}